# A New Hybrid Multi-Objective Scheduling Model for Hierarchical Hub and Flexible Flow Shop Problems

**Sina Aghakhani [1] and Mohammad Sadra Rajabi [2,\***

1. Department of Industrial and Manufacturing Systems Engineering, Iowa State University, Ames, IA, USA;

   sina75@iastate.edu

2. School of Civil Engineering, College of Engineering, University of Tehran, Tehran, Iran;

   msadra.rajabi@gmail.com

\* Correspondence: msadra.rajabi@gmail.com

**Abstract:** Technologies and lifestyles have been increasingly geared toward consumerism in recent years. In general, customers are looking to receive their orders in the fastest time possible and to purchase at a reasonable price. Consequently, the importance of having an optimal delivery time is increasingly evident these days. One of the structures that can meet the demand for large supply chains with numerous orders is the hierarchical integrated hub structure. Such a structure will improve efficiency and reduce chain costs. To make logistics more cost-effective, hub-and-spoke networks are necessary as a means to achieve economies of scale. Many hub network design models only consider hub type but do not take into account the hub scale measured by freight volume. This paper proposes a multi-objective scheduling model for hierarchical hub structures (HHS), which is layered from top to bottom. In the third layer, the central hub takes factory products from decentralized hubs and sends them to other decentralized hubs, to which customers are connected. In the second layer, non-central hubs are responsible for receiving products from the factory and transferring them to central hubs. These hubs are also responsible for receiving products from central hubs and sending them to customers. Lastly, the first layer contains factories responsible for producing products and providing for their customers. The factory uses the Flexible Flow Shop platform and structure to produce its products. The model's objective is to minimize transportation and production costs as well as product arrival times. To validate and evaluate the model, small instances have been solved and analyzed in detail with the weighted sum and ε-constraint methods. Consequently, based on the ideal mean distance (MID) metric, two methods were compared for the designed instances. As NP-hardness causes the previously proposed methods which solve large-scale problems to be time-consuming, a meta-heuristic method was developed to solve the large-scale problem.

## 1. INTRODUCTION

Since the deregulation of the US aviation market in 1978, the network configuration of airlines has been profoundly affected and reconstructed [1]. As a result, a number of trunkline companies reorganized their networks from point-to-point (PP) systems into hub-and-spoke systems. In the mentioned system, the hub serves as a central airport for mapping flight routes, whereas the spokes serve as the flight routes taken from the hub. The transition to hub-and-spoke systems minimizes the logistics portion of the network design cost in aviation systems and other major transportation systems [2]. Because of this cost-effective strategy, the proposed system has been updated and applied to numerous optimization fields [3–6].

Engineering and transportation challenges are being addressed with technological advances, especially in time management, scheduling, and safety concerns [7–12]. As a physical network system, hub and spoke is based on logistics and pertains to the transportation of freight. It involves a hub that moves goods from one spoke to another. The design of hub and spoke networks has been receiving attention in a variety of application fields, including transportation, telecommunications, computer networks, postal delivery, less-than-truck loading, and supply chain management [13,14]. In hierarchical hub problems, demand nodes are routed to facilities known as strategic hubs. Although this is the current process in hierarchical hubs, their designs are constantly evolving due to the emergence of newer multi-structural transportation and distribution networks [15]. The variety of transportation systems adds multimodality to these issues [16]. Because strategic planning is a major component of this field, all possible decisions must be carefully considered for their long-term implications. Therefore, processes are necessary to establish a framework for reliable decisions. Network scheduling and production scheduling are two decision-making processes that decide how to allocate limited resources, like machines, operators, and tools, within a network and how to deliver goods to the desired destination in that network [12,17,18].

The rest of the paper is organized as follows. Section 2 presents the literature review. In Section 3, the problem definition, as well as the mathematical model, has been explained. The solution methods, instances, and investigation of instances have been discussed in Section 4. The discussion and conclusion have been explained in Section 5 and Section 6, respectively.

## 2. LITERATURE REVIEW

### a. Hierarchical hub problems

In hierarchical systems, facilities are either connected from top to bottom or from bottom to top. The lower level is called "the first level," where customers' nodes are generally allocated to, while the higher level is called the "Nth level." Current literature has examined hierarchical hub problems (HHP) from many perspectives, such as location, structure, routing, scheduling, and solution approach. Hub and spoke networks offer the following advantages compared to the complete network. Firstly, the hub network consists of fewer links. The link usage ratio in the building and operating of the network declines, so the overall network utilization increases. Secondly, according to various studies, H&S networks have a discount factor ($\alpha$) due to the economies of scale derived from the inter-hub transport and, consequently, are relatively inexpensive. Nevertheless, H&S networks do not guarantee a more efficient logistic design than an entire network. Consequently, some studies also examine the logistics path of cargo from origin to destination from the perspective of the entire network and reflect direct shipments. In addition, studies have shown that the number of hubs is a decision variable in Hub Location Problem [19–21].

A number of researchers have investigated the location and structure of the hierarchical problem. Schultz (1970) and Calvo and Marks (1973) were the first researchers that studied hierarchical facility location in multiple-layer structures [22,23]. Dokmeci (1973) proposed a three-level approach for solving the hierarchical location problem's best location and scale of facilities [24]. Lastly, Chou (1990) presented multiple allocations in a hierarchical hub location problem (HHLP) of airline networks, and he employed an enumeration approach for this problem [25]. Operations research literature has also provided insight into types of objective approaches regarding the location and structure of HHPs. Yaman (2009) presented a single objective model of the hierarchical hub median location problem (HHMLP), which included three layers: central hubs at the top level, a non-central hub at the second level, and customers' nodes at the third level [3]. Alumur et al. (2012) explored the hierarchical hub location problem with respect to potential candidates for establishment and routes for delivery. They constructed four levels which consisted of the following: the nodes of demand in level zero, the non-central hubs in level one, a median non-central hub in level three, and a central hub in the last level [26]. Davari and Zarandi (2012) investigated three levels of services with a fuzzy demand in hierarchical hub median location problems. They solved their HHMLP by using Yaman's model (2009), a variable neighborhood search (VNS), and CPLEX software [27].

Routing, scheduling, and algorithms are other aspects of HHPs that researchers have evaluated. Okan Dukkanci and Bahar Y. kara. (2017) worked on a scheduling and routing model for the hierarchical hub location problem and solved their model through a heuristic approach, which delivered results in a reasonable time [28]. Melahat Khodemani-Yazdi et al. (2019) produced a multi-objective hierarchical hub location model that minimized hub facilities costs, transportation costs, and route length through these three algorithms: NsgaII, HAS, and game theory variable neighborhood fuzzy invasive weed optimization (GVIWO). The results showed that the GVIWO algorithm generated better results than the other algorithms [29].

Shang et al. (2021) developed a bi-objective hierarchical multimodal hub location problem to simultaneously minimize the overall system-wide costs and the maximum amount of delivery time. The problem is different from the classic hub location problem in the context of designing a hierarchical multimodal hub-and-spoke network that includes multiple modes of transportation, multiple classes of hubs, and corresponding layers [17]. Bhattacharjee and Mukhopadhyay (2021) presented a Multi-Objective version of the Single-Allocation Hub Median Problem with the aim of minimizing the overhead cost associated with hub and central hub nodes and the total communication cost of the network. In a part of the study, Non-dominated Sorting Genetic Algorithm-II is used to solve the problem, as well as classical Genetic Algorithms on each objective respectively [30].

**b. Flexible flow shop**

For the past four decades or so, the flexible flow shop (FFS) scheduling problem has attracted many researchers, and numerous research articles have been published on this topic. This is because the flow shop problem (FSP) is one of the most common scheduling problems in production systems. In the flow shop environment, jobs (productions) should be processed on all machines, while a flexible flow shop problem (FFSP) should consist of sets of production stages so that at least one of the stages has two parallel machines. There is a multitude of issues with flexible flow shop problems, including unrelated parallel machines, release dates, setup times, precedence constraints, eligibility, and batch processing. Additionally, there are issues related to resources with different objective functions, such as makespan, energy consumption, total tardiness, green function, and inventory holding costs. The following literature addresses the various problems of the flexible flow shop problem.

Jenabi et al. (2007) presented an unrelated parallel machines model over a specific planning horizon to minimize setup time and inventory holding costs [31]. Ruiz and Stützle (2008) displayed a model with constraints consisting of the release dates, unrelated parallel machines, machine eligibility, probability of anticipatory and non-anticipatory setup times, precedence constraints, and time lags [32]. Jayamohan and Rajendran (2000) worked on a series of new dispatching methods to reduce performance measure types such as the average, maximum, and variance of tardiness in dynamic problems [33]. Kianfar et al. (2009) demonstrated four dispatching lemmas to minimize the total tardiness and rejection costs [34].

Moreover, Gupta and Tunc (1991) analyzed the two-stage hybrid flow shop scheduling problem with parallel machines only at the second stage to develop heuristic algorithms that minimize the objective function [35]. Bertel and Billaut (2004) presented the multi-stage scheduling problem and conducted a genetic algorithm (GA) to minimize the weighted number of late productions [36]. Yaping Fu et al. (2019) proposed a multi-objective stochastic model to minimize makespan and energy consumption using three algorithms: MOPSO, NSGA2, MOEA/D [37]. Tian-Soon Lee and Ying-Tai Loong (2019) investigated a literature review of scheduling models and solution methods in flexible flow shop problems [38]. Yi Tan et al. (2017). exhibited batch processing machines and unequal release times of jobs in a flexible flow shop model and performed a decomposition method based on the iterative stage and combined it with the neighborhood search approach [39].

There are more examples of research articles confronting the same flexible flow shop problem. Using a hybrid genetic algorithm, Juárez-Pérez et al. (2022) solved the FFS scheduling problem with sequence-dependent setup times in a grid environment [40]. Zhang et al. (2021) developed a hybrid flow shop scheduling problem containing consistent subplots with the aim of optimizing two conflicting objectives simultaneously: the makespan and the total number of subplots [41]. The paper by Wu et al. (2018) considers variable processing times resulting from renewable energy sources in a multi-objective flexible flow shop scheduling problem [42]. Also, a multi-objective optimization algorithm has been proposed by Li et al. (2018) for solving the hybrid flow shop scheduling problem with consideration for setup energy consumptions [43].

### c. Research contribution

To the best of the authors' knowledge, previous studies have focused primarily on location, allocation, and routing problems, different types of objectives, and solutions approach in hierarchical hub location problems (HHLP), while scheduling problems seem less recurrent. In addition, the review of the flexible flow shop problem analysis shows that there are no hybrid models of hierarchical hub location problems and flexible flow shop models (HHLP-FFS) available. To address this gap, this research study develops a hybrid scheduling model for hierarchical hubs and flexible flow shops.

### 3. PROBLEM DEFINITION

Using the hierarchical structure, our study consists of three levels which will be discussed from top to bottom, respectively. The third level is a central hub that receives factories' products from non-central hubs and sends those products to another non-central hub where the customers' nodes have been connected. The second level includes non-central hubs responsible for receiving products from factories and sending them to central hubs. Besides, they receive products from central hubs or factories and then send them to customers. The first level consists of the customers and factories. The customers want to receive products, and the factories produce their products in a Flexible Flow Shop environment. Figure 1. shows the structure of this model.

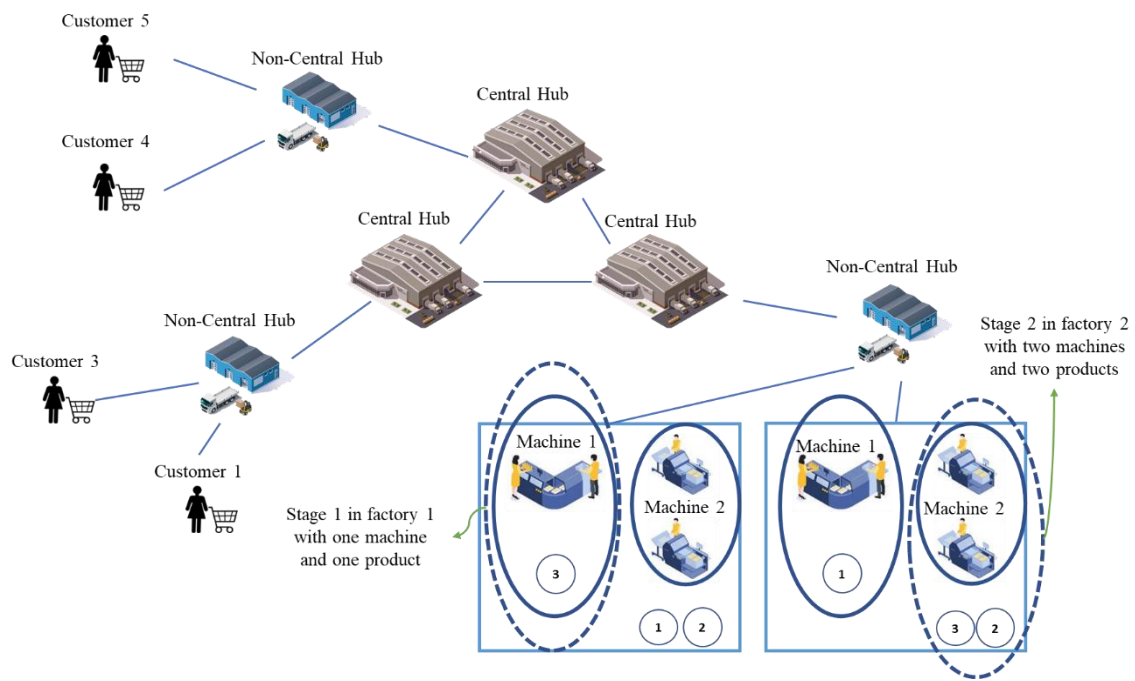

**Figure 1.** The structure of the model

The main structure of the problem has been discussed in the previous section but here; first, we explain the assumptions of the problem and then present the mathematical model based on those defined suppositions and framework. In this model, there are several products, so the least of product types should be more than one. In addition, there are some factories with flexible flow shop environments and customers in a hierarchical hub problem structure. Products process time and the number of machines at each stage are specific. The time spans between nodes (customers or factory), non-central hubs, and central hubs are also clear. There is not any permutation at each stage, and machines are also unrelated. We named non-central hubs connected to customers as NCHC, non-central hubs connected to factories as NCHF, and central hubs as (CH).

**Sets:**

| | |
|---|---|
| $S$ | Set of stages |
| $H$ | Set of non-central hubs connected to customers |
| $P$ | Set of products |
| $F$ | Set of factories |
| $J$ | Set of non-central hubs connected to factories |
| $K$ | Set of the central hubs |
| $C$ | Set of customers |

**Indexes:**

| | |
|---|---|
| $s \in S$ | Index of stages |
| $h \in H$ | Index of non-central hubs connected to customers |
| $p, p' \in P$ | Index of products |
| $f \in F$ | Index of factories |
| $j \in J$ | Index of non-central hubs connected to factories |
| $k, k' \in K$ | Index of the central hubs |
| $c \in C$ | Index of customers |

**Parameters**

$CFR_{fj}^{p}$: Cost of produced product *p* in factory *f* connected to non-central hub *j*.

$c_{jkk'h}$: Cost of sending products from non-central hub *j* (NCHF) to central hub k (CH), then from central hub *k* to another central hub *k'*, and finally from the central hub *k'* to non-central hub *h* (NCHC).

$cc_{jkh}$: Products sending cost from non-central hub *j* to central hub *k* then to non-central hub *h*.

$co_{ch}$: Cost of connecting customer node *c* to non-central hub *h*.

$cg_{hk}$: Cost of connecting non-central hub *h* to central hub *k*.

$cd_{fj}$: Cost of connecting factory *f* to non-central hub *j*.

$cb_{jk}$: Cost of connecting non-central hub *j* to central hub *k*.

$D_{c}^{p}$: The demand of customer *c* for product *p*.

$t_{fj}^{p}$: Period of time between factory *f* and non-central hub *j* for product *p*.

$tt_{jk}^{p}$: Period of time between non-central hub *j* and central hub *k* for product *p*.

$t'^{p}_{k'h}$: Period of time between the central hub *k'* and non-central hub *h* for product *p*.

$ta_{kk'}^{p}$: Period of time between the central hub *k'* and central hub *k* for product *p*.

$V_{fs}$: The number of the machine at stage *s* in factory *f*.

$td_{ch}^{p}$: Period of time between non-central hub *k* and customer *c* for product *p*.

**Binary variables**

$z_{ch}$: If customer *c* connects to the non-center hub *h* is 1, otherwise 0.

$xx_{mps}^{f}$ :      If product $p$ is processed (produced) in factory $f$ by machine $m$ at stage $s$ is 1, otherwise 0.

$zz_{fj}$ :      If factory $f$ is assigned to non-center hub $j$ is 1, otherwise 0.

$x_{jk}$ :      If non-central hub $j$ is assigned to central hub $k$ is 1, otherwise 0.

$z_{ch}$ :      If customer $c$ is assigned to non-central hub $h$ is 1, otherwise 0.

$y_{hk}$ :      If non-central hub $h$ is assigned to customer $k$ is 1, otherwise 0.

$x'^{f}_{mpp's}$ :      If the product $p'$ is processed (produced) immediately after $p$ in factory $f$ on machine $m$ at stage $s$ is 1, otherwise 0.

$v^{p}_{jkk'h}$ :      If the variable $F^{p}_{jkk'h}$ is positive, this variable will be 1, otherwise 0.

$v'^{p}_{jkh}$ :      If the variable $FF^{p}_{jkh}$ is positive, this variable will be 1, otherwise 0.

**Positive variables**

$FR^{p}_{fj}$ :      The number of produced product $p$ in factory $f$ connected to non-central hub j.

$F^{p}_{jkk'h}$ :      The number of product $p$ sent from non-central hub $j$ (NCHF) to central hub $k$ (CH), then from central hub to another central hub $k'$, and finally from the central hub $k'$ to non-central hub $h$ (NCHC).

$FF^{p}_{jkh}$ :      The number of product $p$ sent from non-central hub $j$ (NCHF) to central hub $k$, then from central hub $k$ to non-central hub $h$ (NCHC).

$ST^{p}_{fj}$ :      Maximum arrival time of product $p$ to non-central hub $j$ (NCHF).

$SF^{p}_{h}$ :      Maximum arrival time of product $p$ to non-central hub $h$ (NCHC)

$C_f^{\max}$ :     Maximum completion time of products in factory $f$.

$SG_{ch}^{p}$ :     Maximum arrival time of product $p$ to customer $c$ connected to non-central hub $h$

There are two kinds of objective function in this research. The first one consists of transportation and production costs. Indeed, part a calculates products' sending costs from non-central hub j (NCHF) to central hub k (CH), then from central hub k to another central hub $k'$, and finally from central hub $k'$ to non-central hub h (NCHC). The next part, part b, calculates products' sending costs from non-central hub j to central hub k, then from central hub k to non-center hub h. Part c calculates the cost of connecting customer c to non-central hub h. Part g calculates the production cost of product p in factory f. The cost of connecting non-central hub h to central hub k is calculated in part d. Part e calculates the cost of connecting factory f to non-central hub j. Finally, the cost of connecting non-central hub j to central hub k is calculated in Part f.

$$Min(F1) = \underbrace{\sum_j \sum_k \sum_{k'} \sum_h \sum_p F_{jkk'h}^{p} . c_{jkk'h}^{p}}_{a} + \underbrace{\sum_j \sum_k \sum_h \sum_p FF_{jkh}^{p} . cc_{jkh}^{p}}_{b} + \underbrace{\sum_c \sum_h z_{ch} . co_{ch}}_{c} + \quad (\lambda 1)$$

$$\underbrace{\sum_j \sum_f \sum_p CFR_{fj}^{p} . FR_{fj}^{p}}_{g} + \underbrace{\sum_h \sum_k y_{hk} . cg_{hk}}_{d} + \underbrace{\sum_j \sum_f zz_{fj} . cd_{fj}}_{e} + \underbrace{\sum_j \sum_k x_{jk} . cb_{jk}}_{f}$$

The following objective function minimizes the maximum arrival time of products to customers.

$$Min(F2) = SA \quad (\lambda 2)$$

Constraint (1) assigns factories to non-hub centers. Constraints (2), (3) assign non-hub centers j to hub centers. Each non-central hub should consist of at least one of the factory's nodes. Constraint (3) allocates customers to non-central hub centers. Constraints (5), (6) allocate non-hub centers h to central- hub k so that each central hub should connect to at least one non-central hub.

$$\sum_{j \in J} zz_{fj} = 1 \qquad \forall f \in F \qquad (1)$$

$$\sum_{k \in K} x_{jk} = 1 \qquad \forall j \in J \qquad (2)$$

$$\sum_{j \in J} x_{jk} \geq 1 \qquad \forall k \in K \qquad (3)$$

$$\sum_{c \in C} z_{ch} = 1 \qquad \forall h \in H \qquad (4)$$

$$\sum_{h \in H} y_{hk} \geq 1 \qquad \forall k \in K \qquad (5)$$

$$\sum_{k \in K} y_{hk} = 1 \qquad \forall h \in H \qquad (6)$$

Constraint (7) guarantees that the product p entered to non-central hub h (a kind of non-central hub that only customers are connected to it) should equal to the customer's demand. Constraint (8) determined the produced products in a non-central factory hub (where the factories are only connected to them) based on the products sent to another hub.

$$\sum_{c \in C} D_c^p \cdot z_{ch} = \sum_{j \in J} \sum_{k \in K} \sum_{k' \in K} F_{jkk'h}^p + \sum_{j \in J} \sum_{k \in K} FF_{jkh}^p \qquad \forall h \in H, p \in P \qquad (7)$$

$$\sum_{f \in F} FR_{fj}^p = \sum_{h \in H} \sum_{k \in K} \sum_{\substack{k' \in K \\ k \neq k'}} F_{jkk'h}^p + \sum_{h \in H} \sum_{k \in K} FF_{jkh}^p \qquad \forall j \in J, p \in P \qquad (8)$$

Constraints (9), (10), (11), (12) guarantee that if $x_{jk}$ and $y_{hk}$ are equal to one simultaneously, the variable $F_{jkk'h}^p$ and $FF_{jkh}^p$ will be positive.

$$F_{jkk'h}^p \leq M \cdot x_{jk} \qquad \forall j \in J, k \in K, k' \in K, h \in H, p \in P; k \neq k' \qquad (9)$$

$$F_{jkk'h}^p \leq M \cdot y_{hk} \qquad \forall j \in J, k \in K, k' \in K, h \in H, p \in P; k \neq k' \qquad (10)$$

$$FF^p_{jkh} \leq M.x_{jk} \qquad \forall j \in J, k \in K, h \in H, p \in P \qquad (11)$$

$$FF^p_{jkh} \leq M.y_{hk} \qquad \forall j \in J, k \in K, h \in H, p \in P \qquad (12)$$

Constraint (13) guarantees that if there is a connection between the factory f and non-central hub j, the variable $FR^p_{fj}$ will be positive.

$$FR^p_{fj} \leq M.zz_{fj} \qquad \forall j \in J, f \in F, p \in P \qquad (13)$$

Constraints (14), (15) assign each product to one of the defined machines at each stage depending on production amounts.

$$\sum_{m=1}^{V_{fs}} xx^f_{mps} = 1 \qquad \forall s \in S, f \in F, p \in P \qquad (14)$$

$$\sum_{m=1}^{V_{fs}} xx^f_{mps} \leq M.\sum_{j \in J} FR^p_{fj} \qquad \forall f, p \qquad (15)$$

Constraints (16), (17), (18), (19) determine sequences of products in the factories by machine m at each stage.

$$\sum_{m \in M} \sum_{p \in P} x'^f_{mpps} = V_{fs} \qquad \forall s \in S, f \in F \qquad (16)$$

$$x'^f_{mpps} \leq xx^f_{mps} \qquad \forall s \in S, f \in F, p \in P, m \in M \qquad (17)$$

$$\sum_{p \in P} x'^f_{mpp's} = xx^f_{mp's} \qquad \forall s \in S, f \in F, p' \in P, m \in M \qquad (18)$$

$$\sum_{p \in P; p \neq p'} x'^f_{mp'ps} \leq xx^f_{mp's} \qquad \forall s \in S, f \in F, p \in P, m \in M \qquad (19)$$

Constraints (20),(21) determine variable $v^p_{jkk'h}$. If $F^p_{jkk'h}$ gets a positive value, In other words, if products are in flow among a non-central hub connected to the factory, a non-central hub connected to the customer, and the central hub, then the binary variable $v^p_{jkk'h}$ should be 1.

$$v^p_{jkk'h}.M \geq F^p_{jkk'h} \qquad \forall j \in J, k \in K, k' \in K, h \in H, p \in P \qquad (20)$$

$$v^p_{jkk'h} \leq F^p_{jkk'h} \qquad \forall j \in J, k \in K, k' \in K, h \in H, p \in P \qquad (21)$$

Constraints (22),(23) determine variable $v'^p_{jkh}$. If $FF^p_{jkh}$ gets a positive value, in other words, when products are in flow among the non-central hub connected to the factory (NCHCF), a non-central hub connected to the customer (NCHCC), and the central hub (CH), then the binary variable $v'^p_{jkh}$ should be 1.

$$v'^p_{jkh} \cdot M \geq FF^p_{jkh} \qquad \forall j \in J, k \in K, h \in H, p \in P \qquad (22)$$

$$v'^p_{jkh} \leq FF^p_{jkh} \qquad \forall j \in J, k \in K, h \in H, p \in P \qquad (23)$$

Constraint (24) determines the maximum arrival time of products to NCHCF. Additionally, Constraint (25) determines the maximum arrival time of products from NCHCF to CH, then from CH to another CH Finally from CH to NCHCC.

$$ST^p_j \geq t^p_{fj} + C^{\max}_f - M \cdot (1 - ZZ_{fj}) \qquad \forall j \in J, f \in F, p \in P \qquad (24)$$

$$SF^p_h \geq ST^p_j + tt^p_{jk} + t'^p_{kh} + ta^p_{kk'} - M \cdot (1 - v^p_{jkk'h}) \qquad \begin{array}{l}\forall j \in J, k \in K, k' \in K \\ , h \in H, p \in P\end{array} \qquad (25)$$

Constraint (26) determines the arrival time of products from NCHCF to CH and then from CH to NCHCC. Furthermore, Constraint (27) determines the arrival time of products from NCHCC to the customer.

$$SF^p_h \geq ST^p_j + tt^p_{jk} + t'^p_{kh} - M \cdot (1 - v'^p_{jkh}) \qquad \forall j \in J, k \in K, h \in H, p \in P \qquad (26)$$

$$SG^p_{ch} \geq SF^p_h + td^p_{ch} - M \cdot (1 - z_{ch}) \qquad \forall c \in C, h \in H, p \in P \qquad (27)$$

Time sequences of products in the factory with a flexible flow shop system are calculated Constraints (28), (29), (30).

$$ht^f_{ps} \geq ht^f_{ps-1} + \sum_{j \in J} FR^p_{fj} \cdot pr^f_{ps-1} - (1 - x'^f_{mpp's}) \cdot M \qquad \begin{array}{l}\forall f \in F, s \in S, p \in P \\ , p' \in P'; p = p'\end{array} \qquad (28)$$

$$ht^{f}_{p's} \geq ht^{f}_{ps} + \sum_{j \in J} FR^{p'}_{fj} \cdot pr_{p's-1} - (1 - x'^{f}_{mpp's}).M \qquad \forall f \in F, s \in S, p \in P \\, p' \in P'; p \neq p' \qquad (29)$$

$$ht^{f}_{p's} \geq ht^{f}_{p's-1} + \sum_{j \in J} FR^{p'}_{fj} \cdot pr_{p's} - (1 - x'^{f}_{mpp's}).M \qquad \forall f \in F, s \in S, p \in P \\, p' \in P'; p \neq p' \qquad (30)$$

Constraint (31) calculates the arrival time of products to customers. Besides, Constraint (32) calculates the time that final customers receive their products; in other words, it is the maximum system scheduling time.

$$C^{max}_{f} \geq ht^{f}_{ps} \qquad \forall f \in F, s \in S, p \in P \qquad (31)$$

$$SA \geq SG^{p}_{ch} \qquad \forall c \in C, p \in P, h \in H \qquad (32)$$

## 4. SOLUTION APPROACH

In this section, a small example has been presented and investigated thoroughly to evaluate the proposed multi-objective model. Afterward, ten examples have been provided and solved by weighted sum and ε-constraint methods. In the weighted sum method, the weighted sum of the objective functions is optimized while the efficient solutions are obtained by varying the weights. The problem framework is stated as follows:

$$\text{Max } (w1 \times f1(x) + w2 \times f2(x) + \ldots + wp \times fp(x))$$

Subject to:

$x$ belongs S

In the ε-constraint method, we optimize one of the objective functions using the other objective functions as constraints; the entire structure is shown below:

$$\text{Max } f_1(x)$$

Subject to:

$$f2(x) \geq e2$$

$$f3(x) \geq e3$$

$$\ldots$$

$$fp(x) \geq ep$$

$$x \text{ belongs } S$$

### a. Validation of the model

Here, a small-sized problem has been tested based on Table 1. The proposed model has been validated and investigated using "the weighted sum method" through one point of Pareto front with weights $W_1 = 0.5$, $W_2 = 0.5$ (i.e., $W_1$ is a weight of costs functions and $W_2$ is a weight of time function). This instance is also solved by Gams 24.1.2 on a 2.53 GHz CPU equipped with 6 Gigabytes of RAM.

**Table 1.** The set of small designed instances.

| Notations | Value | Notations | Value |
|---|---|---|---|
| $|S|$ | 2 | $|C|$ | 2 |
| $|P|$ | 2 | $|M|$ | 2 |
| $|H|$ | 2 | $|J|$ | 2 |
| $|F|$ | 2 | $V_{fs}$ | $\begin{bmatrix} 1 & 2 \\ 2 & 1 \end{bmatrix}$ |

**Table 2.** The parameters of designed instances.

| Parameters | Value | Parameters | Value | Parameters | Value |
|---|---|---|---|---|---|
| $co_{ch}$ | $U(50,80)$ | $t^p_{fj}$ | $U(4,6)$ | $cg_{hk}$ | $U(40,50)$ |

| | | | | | | | |
|---|---|---|---|---|---|---|---|
| $c^p_{jkk'h}$ | | $U(30,60)$ | $tt^p_{jk}$ | $round(U(6,9))$ | $cd_{fj}$ | | $U(40,80)$ |
| $cc_{jkh}$ | | $U(40,70)$ | $t'^p_{k'h}$ | $round(U(1,5))$ | $cb_{jk}$ | | $U(30,50)$ |
| $D^p_c$ | | $U(10,20)$ | $ta^p_{kk'}$ | $round(U(3,6))$ | $pr^p_{fs}$ | | $U(1,2)$ |
| $cfr^p_{ch}$ | | $U(40,50)$ | $td^p_{ch}$ | $round(U(2,8))$ | | | |

After solving the model, the responses and interpretation of the small instance are presented as follows:

The factories with indexes 1,2 were connected to the non-central factory hub with index 2, where it is a spoke of the central hub with index 1 (i.e., $zz_{21}=1; zz_{22}=1; x_{12}=1; x_{21}=1$). Moreover, all customers were connected to a non-central customer hub with index 2 while its central hub with is 2 (i.e., $z_{21}=1; z_{22}=1$, $y_{12}=1$, $y_{21}=1$). The obtained structure from the above instance is shown in Figure 2.

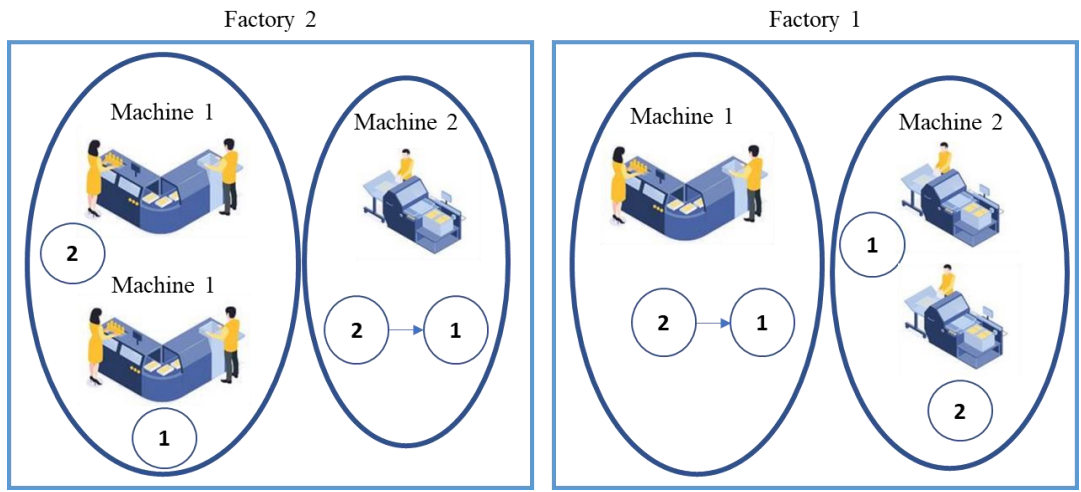

**Figure 2.** The allocation and sequences of products in each stage of the factory

In the following, the amounts of products that should be produced and delivered from non-central factory hub are equal to $FR^1_{12}=4$; $FR^1_{22}=5$; $FR^2_{12}=1$; $FR^2_{22}=10$. Moreover, there is no direct delivery from the non-central factory hub to the central hub then to the non-central customer hub, so these variables are zeros (i.e., $FF^p_{jkh} = 0$), but the variable's value of $F^p_{jkk'h}$ is $F^1_{2121} = 9$ ; $F^2_{2121} = 11$. The products sequences in factory 1,2 and their completion time has also been determined as follows:

$$ht^2_{11} = 1; ht^1_{11} = 9; ht^2_{12} = 3; ht^1_{11} = 17 \quad \text{So} \quad C^{max}_1 = 17.$$

$$ht^2_{21} = 20 ; ht^1_{21} = 10 ; ht^2_{22} = 40 ; ht^1_{22} = 20 \quad \text{So} \quad C^{max}_2 = 40.$$

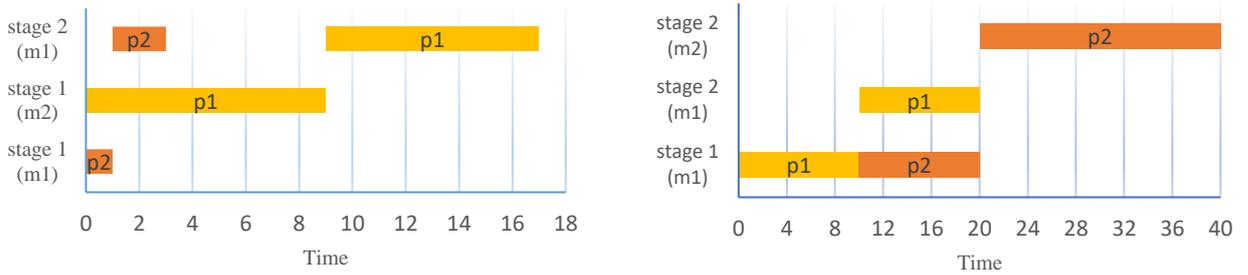

**Figure 3.** The completion time of Factory 1 (left) and Factory 2 (right) Gant chart

Furthermore, the other variables of the shipping time that depend on the variables $F^p_{jkk'h}$ are as below:

$ST_{12} = 44$ ; $ST_{22} = 46$ . $SF_{11} = 57$, $SF_{21} = 59.658$ , $ST_{22} = 46$ . $SG_{111} = 60.01$ ; $SG_{131} = 61.01$ ; $SG_{231} = 61.658$ . $SG_{121} = 62.01$, $SG_{221} = 62.658$ , $SG_{211} = 63.658$.

So, the final maximum completion time (i.e., $SA$) is equal to 63.658. Figure 4. shows the whole network of the generated instance.

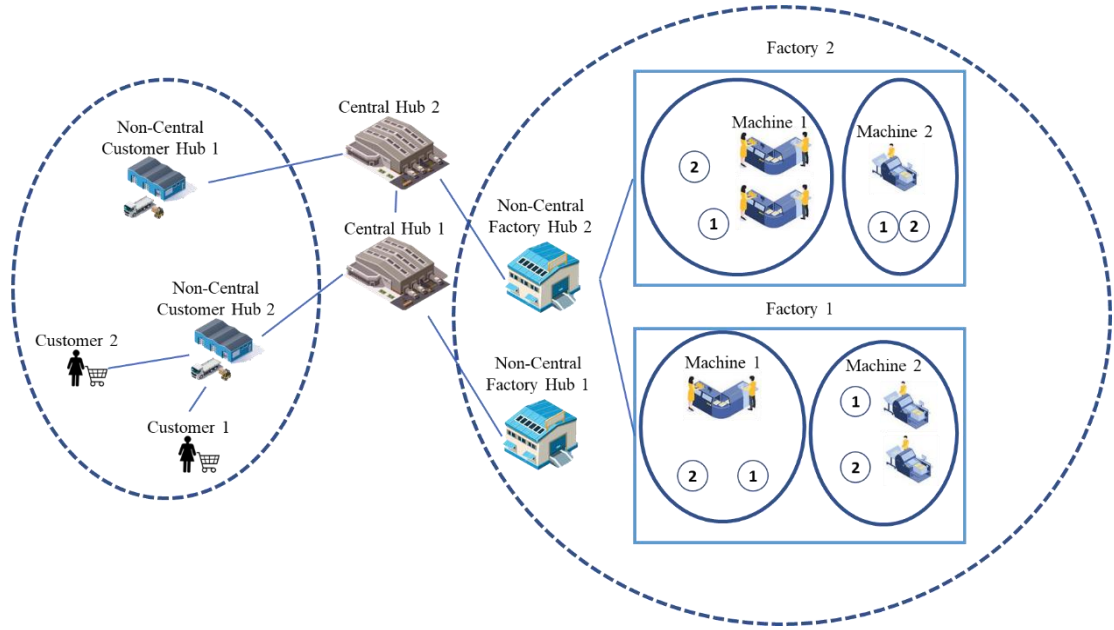

**Figure 4.** The network of the small instance

## b. Numeral experiments

According to the values of variables and the above explanations, the results show the logical relations in terms of structure and arrival time of products. So it assures us to solve and discuss ten instances for further evaluation according to Table 2 and Table 3.

**Table 3.** Designed instances

| Test problem | Sets | | | | | | | Parameter |
|:---:|:---:|:---:|:---:|:---:|:---:|:---:|:---:|:---:|
| | $|H|$ | $|J|$ | $|P|$ | $|K|$ | $|F|$ | $|C|$ | $|S|$ | $V_{fs}$ |
| 1 | 2 | 2 | 2 | 2 | 2 | 2 | 2 | $round(U(1,2))$ |
| 2 | 3 | 2 | 2 | 2 | 3 | 2 | 2 | $round(U(1,2))$ |
| 3 | 2 | 2 | 3 | 2 | 2 | 3 | 2 | $round(U(1,2))$ |
| 4 | 2 | 2 | 3 | 2 | 2 | 3 | 2 | $round(U(1,2))$ |
| 5 | 2 | 2 | 2 | 1 | 3 | 3 | 2 | $round(U(1,2))$ |
| 6 | 2 | 2 | 2 | 2 | 3 | 3 | 2 | $round(U(1,2))$ |
| 7 | 2 | 2 | 2 | 2 | 1 | 2 | 3 | $round(U(1,3))$ |
| 8 | 3 | 2 | 3 | 2 | 1 | 2 | 3 | $round(U(1,3))$ |
| 9 | 2 | 1 | 3 | 1 | 1 | 3 | 4 | $round(U(1,3))$ |
| 10 | 3 | 3 | 2 | 3 | 1 | 4 | 2 | $round(U(1,5))$ |

These instances have been solved by the weighted sum and ε-constraint methods. The Pareto Front of the weighted sum method has been obtained based on Table 4, and the Pareto Front of ε-constraint approach has been determined by eight optional points. The determination of such weights for the weighted sum method in Table 4 depends on managerial decisions and the managers' discretion.

**Table 4.** Weights of objective functions in weighted sum method

| Number of weights | Weight of cost functions(W1) | Weight of time function(W2) |
|---|---|---|
| 1 | 0.1 | 0.9 |
| 2 | 0.2 | 0.8 |
| 3 | 0.091 | 0.909 |
| 4 | 0.0001 | 0.9999 |
| 5 | 0.9 | 0.1 |

The results have been evaluated by the mean ideal distance (MID) metric. This metric is used to calculate the distance between the Pareto solution and an ideal point, point (0,0). The lower the value of this metric for the method, the better the performance and results. The MID metric is determined as below:

$$MID = \frac{\sum_{i=1}^{n}\sqrt{F_{1i}^2 + F_{2i}^2}}{n}$$

Where $i$ is a point of Pareto Front and $n$ is the total points of Pareto Front.

Finally, the results are shown by two methods in Table 5. According to the MID metric, the weighted sum method has the least values, approximately.

**Table 5.** The results of instances are based on the mid metrics.

| Number of instance | MID | |
| --- | --- | --- |
| | weighted sum method | ε-constraint |
| 1 | 1366.04 | 1456.67 |
| 2 | 2392.25 | 2793.64 |
| 3 | 2264.26 | 2393.91 |
| 4 | 2265.29 | 2395.61 |
| 5 | 2254.99 | 2279.10 |
| 6 | 2178.88 | 2261.40 |
| 7 | 1471.04 | 1547.63 |
| 8 | 1525.28 | 1737.77 |
| 9 | 2623.40 | 2670.86 |
| 10 | 2794.20 | 2942.09 |

Figure 5. shows the comparison of two methods based on the MID metric. One of the reasons that the MID metric of the ε-constraint approach is more than the weighted sum method can be managerial decisions and the managers' discretion. Therefore, if we change these weights, the Pareto front of this method may change in terms of the MID metric. The second reason that affects the quality of MID is the initial number of considered Pareto front points. Finally, we cannot claim which methods are better because it depends on the number of Pareto fronts, weights, epsilon dataset, and nature of indexes. We can only mention that for these instances, the MID metric of the weighted sum method is less than the ε-constraint.

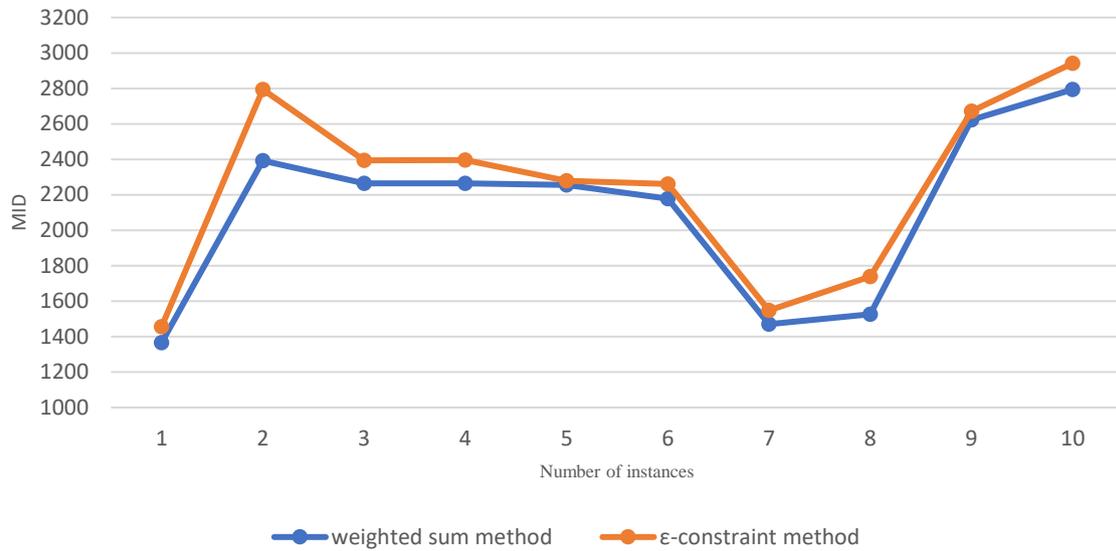

**Figure 5.** Comparison of ε-constraint and weighted sum methods

## 5. DISCUSSION

This paper developed a hybrid scheduling model for hierarchical hubs and flexible flow shops. Although Juárez-Pérez et al. [40], Zhang et al. [41], and Wu et al. [42] tried to solve the FFS scheduling problem with different approaches, none of them solved the scheduling problem using HHLP and FFS. It is the first time to hybridize a hierarchical hub problem with an FFS problem and create a multi-objective problem considering cost minimization and delivery time reduction. It should be noted that two of the known methods (i.e., ε-constraint and weighted sum methods) to solve this problem are presented and compared in order not only to analyze and evaluate the model of the problem but also to examine the speed and accuracy of the performance of these two algorithms in solving the problem in small scale. The problem has been formulated as a new mixed-integer linear programming model (MILP) to minimize transportation, production costs, and product arrival times. The analysis results show that the proposed method not only is able to be applied in solving small-scale problems but also has the potential to be applied to large-scale problems as well.

## 6. MANAGERIAL INSIGHTS AND PRACTICAL IMPLICATIONS

One of the most important practical points of this article, which has been mentioned and worked on in a few articles, is to consider the combination of issues related to the delivery of goods through the hierarchical hub model and the problem of scheduling production within the factory. In fact, this model helps managers and business owners not only to use the hierarchical hub system to deliver products to customers but also to use a combination of the hierarchical hub method and flexible flowshop to plan optimally for their production and delivery department. Also, the minimization of the chain cost and the delivery time of products to the customer, which are two of the most important goals of every manufacturer for the production and delivery of products, have been considered in this issue. Considering the chain's hierarchical structure and central and non-central hubs to manage product delivery and flexible flowshop scheduling, business owners can use this method to optimize their chain and network according to their needs. to use

## 7. CONCLUSION

Hierarchical Hub problems involve the establishment of strategic hub facilities and the allocation of demand nodes to them. Meanwhile, with the emergence of new transport and distribution networks with multi-level structures, the design of such networks has evolved. The diversity of transportation systems has added multimodality to these problems. Considering the strategic nature and long-term implications of decision-making in this field, the decisions shall be of high reliability. Production and network scheduling is a decision-making process to allocate limited resources, such as machines, material handling equipment, operators, and tools, to tasks or jobs and deliver the products to the destinations through the network to achieve certain objectives. Combining these two problems and developing an integrated optimization model with the objective of minimizing cost and timespan is a topic that has rarely been worked on in the literature. In the present paper, a new hybrid multi-objective scheduling model for two combined problems, Hierarchical Hub Problem (HHP) and the Flexible Flow Shop problem (FFSP), was developed. The problem has been formulated as a new mixed-integer linear programming model (MILP) to minimize the transportation and production costs and product arrival times. To solve, validate, and evaluate the presented model, the weighted sum and ε-constraint methods for small-scale problems through Gams software 24.1.2 using CPLEX solver have been utilized. Furthermore, the mean ideal distance (MID) metric was used to compare these two methods. Based on the comparison, the weighted sum method performs better than the ε-constraint method.